# Dynamics on the SU(2) fundamental domain

Bas van den Heuvel* and Pierre van Baal

Instituut-Lorentz for Theoretical Physics, University of Leiden,
PO Box 9506, NL-2300 RA Leiden, The Netherlands.

For SU(2) gauge theory on the three-sphere we focus on a subspace of modes of the gauge field that contains the tunnelling paths and the sphalerons and on which the energy functional is degenerate to second order in the fields. The ultimate goal is to study the $\theta$-dependence of the low-lying states for this model by imposing boundary conditions on the fundamental domain.

## 1. INTRODUCTION

It is our aim to study the dynamics of the low-energy modes of pure SU(2) gauge theory defined in a finite volume [1–3]. Taking the finite volume to be small results, through the mechanism of asymptotic freedom, in a small coupling constant. Gradually increasing the volume then allows us to monitor the onset of non-perturbative phenomena. Especially when the non-perturbative effects manifest themselves appreciably only in a small number of low-lying energy modes, a hamiltonian formulation is useful.

We are interested in the influence of the multiple vacuum structure of the theory on the glueball spectrum; in particular we would like to see the dependence of the energies on the $\theta$-angle. Our strategy will be as follows. We will treat the high-energy modes of the field gaussian, that is, we assume the potential to be quadratic and take harmonic oscillator wave functions. For the low-energy modes, the onset of non-perturbative effects means that the potential will start to deviate from quadratic behaviour. This means that we have to replace the harmonic oscillator wave functions by other functions. The $\theta$-dependence will emerge through boundary conditions on these functions.

The dynamics of these modes can be regarded as an effective low-energy theory. Similar to a Born-Oppenheimer approximation we integrate out the fast (high-energy) modes and are left with an effective theory of the slow modes. If we are at energies at which the higher modes cannot be excited, these modes will, through virtual processes, in first order only result in a renormalisation of the coupling constant.

Before we can hope to tackle this quantum mechanical problem, we have to have a good understanding of the physical configuration space of our system. Since we are dealing with a non-abelian gauge theory, this is a non-trivial problem in its own.

To make contact with lattice calculations it would be most natural to take the finite (spatial) volume to be a 3-torus $T^3$. Detailed knowledge of the vacuum structure in this case is however limited. In particular the instantons, which are the gauge field configurations that describe tunnelling between different vacua, are not known exactly. To circumvent this problem we take our space to be the three-sphere $S^3$ [4]. The instantons in the space-time $\mathbb{R} \times S^3$ can be obtained from those in $\mathbb{R}^4$ by a conformal mapping. The technical set-up for this analysis can be found in [2,3].

## 2. THE PHYSICAL CONFIGURATION SPACE

Let $\mathcal{A}$ be the set of all gauge fields $A_i : S^3 \to su(2)$. We are working in a hamiltonian formalism, so we set $A_0 = 0$. Let $\mathcal{G}$ be the set of all time independent gauge transformations $g : S^3 \to SU(2)$. These act in the standard way:

$$\left({}^g A\right)_i = g^{-1} A_i g + g^{-1} \partial_i g = g^{-1} D_i^{(A)} g. \qquad (1)$$

---

The *physical configuration space* is the space of gauge orbits $\mathcal{A}/\mathcal{G}$. We would like to have a *fundamental domain*, that is, a set of gauge fields which is in one-to-one correspondence with the physical configuration space.

As a representative of an orbit, we choose the gauge field that has lowest norm [5,6]. The norm comes from the standard inner product $\langle A, B \rangle = \int_{S^3} \mathrm{tr}(A^\dagger B)$. This leads to the following candidate for the fundamental domain:

$$\Lambda \equiv \{A \in \mathcal{A} \mid ||{}^g A|| \geq ||A|| \quad \forall g \in \mathcal{G}\}. \quad (2)$$

One can show that $A \in \Lambda$ implies that $A$ is transversal ($\partial_i A_i = 0$) and that the Faddeev-Popov operator is positive ($FP(A) = -\partial_i D_i \geq 0$). The set of fields for which $FP \geq 0$ is called the Gribov region $\Omega$. Both $\Omega$ and $\Lambda$ are convex subsets of the vector space $\Gamma$ consisting of transversal fields. Points where the boundary of $\Omega$ (the Gribov horizon) touches $\Lambda$ are called singular boundary points. Doing the dynamics on $\Lambda$, these are the only points where the coordinate singularity associated with the Gribov horizon shows up.

If the minimum along the gauge orbit is degenerate ($||{}^g A|| = ||A||$ for some $g$), we have to identify these configurations to obtain the one-to-one correspondence $\Lambda \cong \mathcal{A}/\mathcal{G}$. These boundary identifications give $\Lambda$ the required non-trivial topology [7]. We have depicted the various concepts defined here in Fig. 1.

## 3. THE EFFECTIVE THEORY

To find the spectrum of the low-energy theory, we basically have to solve the eigenvalue problem

$$\mathcal{H}\psi = \left(-\frac{1}{2}\frac{\delta^2}{\delta A^2} + V(A)\right)\psi = E\psi. \quad (3)$$

For our model the space of (perturbative) lowest energy modes is 18-dimensional. This is the lowest eigenspace of the quadratic part of the potential $V$. We now have to restrict ourselves to the cross section of this space with the fundamental domain and then solve for the lowest eigenfunctions of the hamiltonian. To make this problem well-defined we have to specify boundary conditions. To keep things transparent we now focus on a subspace of the 18-dimensional space.

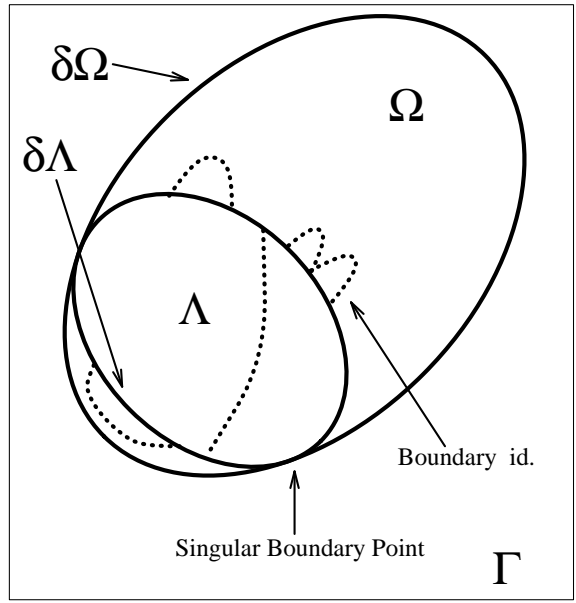

Figure 1. The relevant subsets of gauge fields.

This two-dimensional space of field configurations (Fig. 2) contains three copies of the vacuum (large dots), close to each of which the potential rises quadratically. This space is important because of all the tunnelling paths connecting the vacua, it contains those paths that have the lowest energy barrier. These barrier configurations are saddle points of $V$ and are called sphalerons (small dots); they are also gauge copies of each other.

In this $(u,v)$ plane we can calculate $\Omega$ and a lower bound $\tilde{\Lambda}$ on the fundamental domain: $\tilde{\Lambda} \subset \Lambda$. The lower bound is obtained by defining a Faddeev-Popov-like operator in the fundamental representation. We then have

$$||{}^g A||^2 - ||A||^2 = \langle g, FP_{1/2}(A)g \rangle \geq \mu_1 ||g||^2, \quad (4)$$

with $\mu_1$ the lowest eigenvalue of $FP_{1/2}$. Using the inclusion $\tilde{\Lambda} \subset \Lambda \subset \Omega$, we explicitly found singular boundary points.

We are interested in energies which are high enough to see deviations from the perturbative behaviour, but low enough to restrict these deviations to the lowest modes. This means that we are at energies of the order of the sphaleron energy: here the wave functional starts to spread out over configuration space. Note that WKB methods are no longer applicable here.

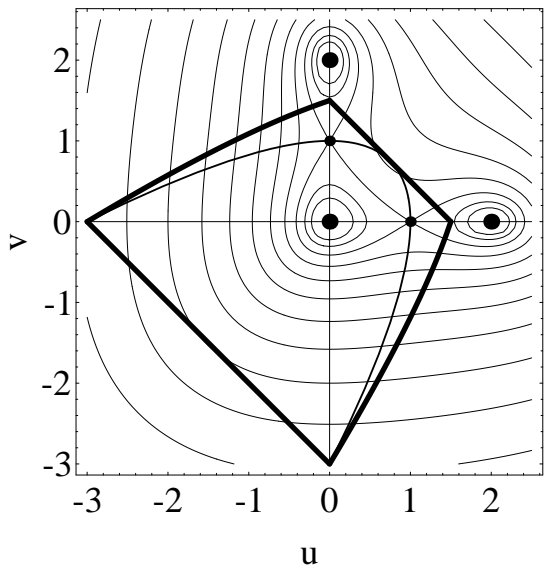

Figure 2. Location of the classical vacua, sphalerons, lines of equal potential, the Gribov horizon (fat sections) and the lower bound $\tilde{\Lambda}$ for the fundamental domain: $\tilde{\Lambda} \subset \Lambda \subset \Omega$.

At these energies the only relevant boundary identification is the one at the sphalerons. At other boundary points the potential will be (much) higher then the energy of the wave functional. This means that the wave functional will have exponentially decayed at these points and that the precise boundary conditions will not have a big influence on the spectrum. By the same token, the precise location of the boundary in these regions is not important either. This gives us the freedom to choose tractable boundary conditions.

At the sphalerons however, the boundary conditions are fixed. Since the gauge transformation connecting the two sphalerons has winding number one, we have to set

$$\psi(A(1,0)) = e^{i\theta}\psi(A(0,1)), \qquad (5)$$

thus introducing the $\theta$-angle.

## 4. CURRENT WORK

We have constructed a boundary with suitable boundary conditions for the full 18-dimensional problem. Since all these modes are perturbatively degenerate, they have to be treated on the same footing. The 18 modes naturally decompose in $9 + 9$ modes, where each set contains one of the sphalerons. The idea is to use polar coordinates in each sector and impose the boundary conditions in the two radial variables. A complication here is that the directions orthogonal to the tunnelling paths at the sphalerons have to be mapped to each other under the boundary conditions in the proper way.

Next we have constructed a basis of functions that incorporate these boundary conditions. Here the rotational symmetry and the residual gauge symmetry (associated to constant gauge transformations) have to be taken into account properly. Once this has been done we can perform a Rayleigh-Ritz analysis to find the spectrum. The next step would be to incorporate quantum corrections coming from the other modes. We hope to elaborate on these matters in the near future.

## REFERENCES

1. J. Koller and P. van Baal, Nucl. Phys. B302 (1988) 1; P. van Baal, Acta Phys. Pol. B20 (1989) 295.
2. P. van Baal and N.D. Hari Dass, Nucl. Phys. B385 (1992) 185.
3. P. van Baal and B.M. van den Heuvel, Nucl. Phys. B417 (1994) 215.
4. R.E. Cutkosky and K. Wang, Phys. Rev. D37 (1988) 3024.
5. M.A. Semenov-Tyan-Shanskii and V.A. Franke, J. Sov. Math. 34 (1986) 1999.
6. D. Zwanziger, Nucl. Phys. B412 (1994) 657 and references therein.
7. I. Singer, Comm. Math. Phys. 60 (1978) 7.
8. P. van Baal and R.E. Cutkosky, Int. J. Mod. Phys. A (Proc. Suppl.) 3A (1993) 323.